\documentclass [12pt,eqsecnum,amsfonts,amssymb,aps]{revtex4}

\input epsf
\usepackage{graphicx}

\newcommand{\beq}{\begin{equation}}
\newcommand{\eeq}{\end{equation}}
\newcommand{\beqs}{\begin{eqnarray}}
\newcommand{\eeqs}{\end{eqnarray}}

\newcommand{\gsim}{\mathrel{\raisebox{-.6ex}{$\stackrel{\textstyle>}{\sim}$}}}

\begin{document}

\baselineskip 6.0mm

\title{Measures of Spin Ordering in the Potts Model with a Generalized 
External Magnetic Field} 

\bigskip

\author{Shu-Chiuan Chang$^{a}$}
\email{scchang@mail.ncku.edu.tw; orcid: 0000-0003-0847-9354}

\author{Robert Shrock$^{b}$}
\email{robert.shrock@stonybrook.edu; orcid:  0000-0001-6541-0893}

\affiliation{(a) \ Department of Physics \\
National Cheng Kung University \\
Tainan 70101, Taiwan}

\affiliation{(b) \ C. N. Yang Institute for Theoretical Physics and \\
Department of Physics and Astronomy \\
Stony Brook University  \\
Stony Brook, N. Y. 11794 \\
USA}

\begin{abstract}

  We formulate measures of spin ordering in the $q$-state ferromagnetic Potts
  model in a generalized external magnetic field that favors or disfavors spin
  values in a subset $I_s = \{1,...,s\}$ of the total set of $q$ values.  The
  results are contrasted with the corresponding measures of spin ordering in
  the case of a conventional external magnetic field that favors or disfavors a
  single spin value out of total set of $q$ values.  Some illustrative
  calculations are included.

\end{abstract}

\maketitle


\pagestyle{plain}
\pagenumbering{arabic}

\section{Introduction}
\label{intro_section}

The $q$-state Potts model \cite{potts} has long been of interest as a
classical spin model in which each spin can take on any of $q$ values
in the interval $I_q = \{1,2,...,q\}$, with a Kronecker delta function
spin-spin interaction between spins on adjacent sites
\cite{wurev,baxterbook}.  In contrast to the $q=2$ case,
which is equivalent to the Ising model, for $q \ge 3$, there are
several different ways that one can incorporate the effect of a
symmetry-breaking (uniform) external magnetic field. The conventional
way is to define this field as favoring one particular spin value out
of the $q$ possible values in the set $I_q$, e.g., \cite{wu78}.  In
\cite{ph,phs,phs2,zhlad} we defined and studied properties of the
$q$-state Potts model with a generalized external magnetic field that
favors or disfavors a subset consisting of more than just one value in
$I_q$.  By convention, with no loss of generality, we take this subset
to consist of the first $s$ values, denoted as the interval $I_s =
\{1,...,s\}$.  The orthogonal subset in $I_q$ is denoted $I_s^\perp =
\{s+1,...,q\}$. In the case that we considered, the value of the 
magnetic field is a constant, consistent with its property as being
applied externally.  More general models with magnetic-like variables
whose field values depend on the vertices have also been discussed
\cite{noble_welsh,sokal_bounds,ellis_moffatt_magnetic}, but we willl not need
this generality here. 

In the present paper we continue the study of the $q$-state Potts model in this
generalized uniform external magnetic field. We discuss measures of spin
ordering in the presence of the external field and formulate an order parameter
for this model. The results are contrasted with the corresponding measures of
spin ordering in the case of a conventional external magnetic field that favors
or disfavors a single spin value in $I_q$. 


\section{Definition and Basic Properties of the Potts Model in a 
Generalized Magnetic Field}
\label{gen_section}

In this section we review the definition and basic properties of the
model that we study. We will consider the Potts model on a graph $G
(V,E)$ defined by its set $V$ of vertices (site) and its set $E$ of
edges (bonds).  For many physical applications, one usually takes $G$
to be a regular $d$-dimensional lattice, but we retain the general
formalism of graph theory here for later use.
In thermal equilibrium at temperature $T$, the
partition function for the $q$-state Potts model on the graph $G$ in a
generalized magnetic field is given by
\beq
Z = \sum_{ \{ \sigma_i \} } e^{-\beta {\cal H}} \ , 
\label{z}
\eeq
with the Hamiltonian
\beq
{\cal H} = -J \sum_{e_{ij}} \delta_{\sigma_i, \sigma_j}
- \sum_{p=1}^q H_p \sum_\ell \delta_{\sigma_\ell,p} \ ,
\label{ham}
\eeq
where $i, \ j, \ \ell$ label vertices of $G$; $\sigma_i$ are
classical spin variables on these vertices, taking values in the set
$I_q = \{1,...,q\}$; $\beta = (k_BT)^{-1}$; $e_{ij}$ is the edge
(bonds) joining vertices $i$ and $j$; $J$ is the spin-spin
interaction constant; and 
\beq
H_p = \cases{ H & if $p \in I_s$ \cr
              0 & if $p \in I_s^\perp$} \ .
\label{hp}
\eeq
Unless otherwise stated, we restrict our
discussion to the ferromagnetic ($J > 0$) version of the model, since the
antiferromagnetic model in a (uniform) external field entails complications due
to competing interactions and frustration.  If $H > 0$, the external field
favorably weights spin values in the interval $I_s$, while if $H<0$, this field
favorably weights spin values in the orthogonal interval $I_s^\perp$. This
model thus generalizes a conventional magnetic field, which would favor or
disfavor one particular spin value.  The zero-field Potts model Hamiltonian
${\cal H}$ and partition function $Z$ are invariant under the global
transformation in which $\sigma_i \to g \sigma_i \ \ \forall \ \ i \in V$, with
$g \in {\cal S}_q$, where ${\cal S}_q$ is the symmetric (= permutation) group
on $q$ objects.  In the presence of the generalized external field defined in
Eq. (\ref{hp}), this symmetry group of ${\cal H}$ and $Z$ is reduced from $S_q$
at $H=0$ to the tensor product
\beq
{\cal S}_s \otimes {\cal S}_{q-s} \ .
\label{symgroup}
\eeq
This simplifies to the conventional situation in which the external magnetic
field favors or disfavors only a single spin value if $s=1$ or $s=q-1$, in
which case the right-hand side of Eq. (\ref{symgroup}) is ${\cal S}_{q-1}$.

We use the notation
\beq
K = \beta J \ , \quad h = \beta H \ , \quad y = e^K \ , \quad v = y-1 \ ,
\quad w=e^h \ .
\label{variables}
\eeq
The physical ranges of $v$ are $v \ge 0$ for the Potts ferromagnet, and $-1 \le
v \le 0$ for the Potts antiferromagnet.  For fixed $J$ and $H$, as $T \to
\infty$, $v \to 0$ and $w \to 1$, while for $T \to 0$ (with our ferromagnetic 
choice $J > 0$), $v \to \infty$; and $w \to \infty$ if $H > 0$ while 
$w \to 0$ if $H < 0$.  Recall that for $q=2$, the 
equivalence with the Ising model with standard Hamiltonian
(denoted with $Is$) 
\beq
{\cal H}_{Is} = -J_{Is} \sum_{e_{ij}} 
\sigma^{(Is)}_i \sigma^{(Is)}_j - H_{Is}\sum_{i} \sigma^{(Is)}_i \ , 
\label{ham_ising}
\eeq
where $\sigma^{(Is)}_i = \pm 1$ makes use of the relations 
$J=2J_{Is}$ and $H=2H_{Is}$.

One can express the Potts model partition function on a graph $G$ in a
form that does not make any explicit reference to the spins $\sigma_i$
or the summation over spin configurations in (\ref{z}), but instead is
expressed in a purely graph-theoretic manner, as a sum of terms
arising from the spanning subgraphs $G' \subseteq G$, where
$G'=(V,E')$ with $E' \subseteq E$.  In zero field, this was done in
\cite{fk}, with the result
\beq
Z(G,q,v) = \sum_{G' \subseteq G} v^{e(G')} \ q^{k(G')} \ . 
\label{fk}
\eeq
For the model with a conventional external magnetic field that favors
or disfavors a single spin value in the set $I_q$, a spanning-subgraph
formula for the partition function was given in \cite{wu78}.  For the
model with a generalized magnetic field that favors or disfavors a
larger set $I_s$ consisting of two or more spin values in the set
$I_q$ and has a value $H_{i,p}=H_p$ that is the same for all vertices
$i \in V$, a spanning-subgraph formula for the partition function was
presented in Ref. \cite{ph} (see also \cite{phs}) and is as follows.
Given a graph $G=(V,E)$, the numbers of vertices, edges, and connected
components of $G$ are denoted, respectively, by $n(G) \equiv n$,
$e(G)$, and $k(G)$. The purely graph-theoretic expression of the
partition function of the Potts model in a generalized magnetic field
in this case is \cite{ph}
\beq
Z(G,q,s,v,w) = \sum_{G' \subseteq G} v^{e(G')} \
\prod_{i=1}^{k(G')} \, (q-s+sw^{n(G'_i)})  \ , 
\label{clusterws}
\eeq
where $G'_i$, $1 \le i \le k(G')$ denotes one of the $k(G')$ connected
components in a spanning subgraph $G'$ of $G$. The formula
(\ref{clusterws}) shows that $Z$ is a polynomial in the variables $q$,
$s$, $v$, and $w$, hence our notation $Z(G,q,s,v,w)$. For the case
where the magnetic field favors (or disfavors) only a single spin
value, i.e., for the case $s=1$, the formula (\ref{clusterws}) reduces
to the spanning subgraph formula for $Z$ given in \cite{wu78} (see
also \cite{hl}). Parenthetically, we mention further generalizations
that are different from the one we study here.  First, one can let the
spin-spin exchange constants $J$ be edge-dependent, denoted as
$J_{ij}$ on the edge $e_{ij}$ joining vertices $i$ and $j$.  Second,
one can let the value of the magnetic-field-type variable be different
for different vertices $\ell \in V$, so $H_p$ is replaced by
$H_{\ell,p}$. With these generalizations, a spanning-subgraph formula
for the partition function was given in \cite{sokal_bounds} and
studied further in \cite{ellis_moffatt_magnetic}.

Focusing on the term $w^{n(G'_i)}$ in (\ref{clusterws}) and letting
$\ell = n(G'_i)$ for compact notation, one can use the factorization relation
\beq
w^\ell-1=(w-1)\sum_{j=0}^{\ell-1}w^j
\label{wellfactorization}
\eeq
to deduce that the variable $s$ enters in $Z(G,q,s,v,w)$, only in the
combination
\beq
t=s(w-1) \ .
\label{tvar}
\eeq
Hence, the special case of zero external field, $H=0$, i.e., $w=1$, is 
equivalent to the formal value $s=0$ (outside the interval $I_s$).

Several relevant identities were derived in \cite{ph,phs}, including 
\beq
Z(G,q,s,v,1) =  Z(G,q,v) \ , 
\label{zw1}
\eeq
\beq
Z(G,q,s,v,0) = Z(G,q-s,v) \ , 
\label{zw0}
\eeq
\beq
Z(G,q,q,v,w) = w^n \, Z(G,q,v) \ ,  
\label{zwsq}
\eeq
and
\beq
Z(G,q,s,v,w) = w^n Z(G,q,q-s,v,w^{-1}) \ . 
\label{zsym}
\eeq
The identity (\ref{zsym}) establishes a relation between the model with $H > 0$
and hence $w > 1$, and the model with $H < 0$ and hence $0 \le w < 1$.  Given
this identity, one may, with no loss of generality, restrict to $H \ge 0$,
i.e., $w \ge 1$, and we will do this below, unless otherwise indicated.

In the limit $n(G) \to \infty$, the reduced, dimensionless free energy per
vertex is 
\beq
f(\{ G \}, q,s,v,w) = \lim_{n(G) \to \infty} \frac{1}{n(G)} \ln[Z(G,q,s,v,w)] 
\ , 
\label{f}
\eeq
where the symbol $\{ G \}$ denotes the formal $n \to \infty$ limit of
a given family of graphs, such as a regular lattice with some
specified boundary conditions. The actual Gibbs free energy per site
is $F(T,H) = -k_BT f(T,H)$.  For technical simplicity, unless
otherwise indicated, we will restrict to the ferromagnetic case
$J > 0$ here;
in \cite{ph,phs,phs2,zhlad} we have also discussed the
antiferromagnetic case. The zero-temperature limit of the
antiferromagnetic version defines a weighted-set chromatic polynomial
that counts the number of assignments from $q$ colors to the vertices
of $G$ subject to the condition that no two adjacent vertices have the
same color, with prefered (disprefered) weighting given to colors in
$I_s$ for $H > 0$ ($H < 0$, respectively).  Here and below, in order
to avoid cumbersome notation, we will use the same symbol $Z$ with
different sets of arguments to refer to the full model, as
$Z(G,q,s,v,w)$ and the zero-field special case, $Z(G,q,v)$.

The partition function of the zero-field Potts model is equivalent to an
important function in mathematical graph theory, namely the Tutte (also called
Tutte-Whitney) polynomial $T(G,x,y)$ \cite{tutte}.  This is defined as
\beq
T(G,x,y) = \sum_{G' \subseteq G} (x-1)^{k(G')-k(G)} (y-1)^{c(G')} \ ,
\label{t}
\eeq
where $c(G')$ denotes the number of linearly independent cycles on $G'$.
Note that $c(G')=e(G')+k(G')-n(G')=e(G')+k(G')-n(G)$. The equivalence
relation is
\beq
Z(G,q,v) = (x-1)^{k(G)}(y-1)^{n(G)}T(G,x,y) \ ,
\label{ztrel}
\eeq
with
\beq
x = 1 + \frac{q}{v} \ , \quad y = v+1 \ ,
\label{xyqv}
\eeq
so that $q=(x-1)(y-1)$. Reviews of the Tutte polynomial and
generalizations include
\cite{noble_welsh,sokal_bounds,ellis_moffatt_magnetic}, 
\cite{bollobas}-\cite{ellis_moffatt_book}. 


\section{Magnetic Order Parameter in the Ising, O($N$), and Potts Model with
a Conventional Magnetic Field} 
\label{conventional_section}

\subsection{Ising and O($N$) Models}

In the Ising model (\ref{ham_ising}) the magnetization per site is given by 
\beq
M(H) = -\frac{\partial F}{\partial H} = \frac{\partial f}{\partial h} \ , 
\label{m_ising}
\eeq
and the spontaneous magnetization is $M \equiv \lim_{H \to 0}M(H)$. This 
$M$ is (i) identically zero in the high-temperature phase where the theory is 
invariant under the global 
${\mathbb Z}_2 \approx S_2$ symmetry and, (ii) for a regular lattice of
dimensionality above the lower critical dimensionality $d_\ell=1$, $M$ is
nonzero in the low-temperature phase where there is spontaneous breaking of 
this global ${\cal Z}_2$ symmetry, increasing from 0 to a maximum of 1 as 
$T$ decreases from the critical temperature $T_c$ to $T=0$.  Alternatively, in 
the ferromagnetic Ising model (\ref{ham_ising}) on a regular lattice, the
(square of the) magnetization can be calculated in the absence of an 
external field as the limit
\beq
M^2 = \lim_{r \to \infty} \langle \sigma^{(Is)}_i \sigma^{(Is)}_j \rangle \ , 
\label{msq_ising}
\eeq
where $\langle \sigma^{(Is)}_i \sigma^{(Is)}_j \rangle$ is the
two-spin correlation function and $r$ denotes the Euclidean distance
between the lattice sites $i$ and $j$ on the lattice. For $T>T_c$,
this correlation function $\langle \sigma^{(Is)}_i \sigma^{(Is)}_j
\rangle \to 0$ as $r \to \infty$, so that $M=0$. Regarding the Ising
model as the $N=1$ special case of an O($N$) spin model, similar
comments hold.  Thus, consider the partition function
\beq
Z_{{\rm O}(N)} = \int \prod_i d\Omega_i e^{-\beta {\cal H}_{{\rm O}(N)}} \ ,
\label{zon}
\eeq
with
\beq
{\cal H}_{{\rm O}(N)} = -J\sum_{\langle ij \rangle} \vec S_i \cdot \vec S_j 
-\vec H \cdot \sum_i \vec S_i \ , 
\label{hamon}
\eeq
where $\vec S_i$ is an $N$-component unit-normalized classical spin at site $i$
on a given lattice and $d\Omega_i$ denotes the O($N$) integration measure. For
zero external field, this model has a global O($N$) invariance.  The presence
of an external magnetic field $\vec H$ explicitly breaks the O($N$) symmetry
down to O($N-1$). For general $\vec H$, one has, for the thermal average of
$\vec M = \vec M(\vec H)$,
\beq
\vec M(\vec H) = -\frac{\partial F}{\partial {\vec H}} \ , 
\label{mon}
\eeq
and the relation 
\beq
|\vec M|^2 = \lim_{r \to \infty} \langle \vec S_i \cdot \vec S_j \rangle \ .
\label{msqon}
\eeq
As usual, the spontaneous magnetization for the Ising and O($N$) models is 
defined as $M_0 = \lim_{H \to 0}M(H)$ and 
${\vec M}_0 = \lim_{|\vec H| \to 0} { \vec M}(\vec H)$, respectively. 


\subsection{Measures of Spin Ordering in Potts Model with Conventional 
Magnetic Field}

The situation is different in the Potts model, even with a conventional
external magnetic field that favors only one spin value. In our formalism, this
means that $I_s$ consists of the single value $s=1$.  Before proceeding to
derive our new results, we review this situation for this conventional case.
For this purpose, it is useful to analyze the properties of the spin-spin
correlation function.  It will be sufficient here and below to assume that the
graph $G$ is a regular $d$-dimensional lattice. Let us denote as $P_{aa}(i,j)$
the probability (in the thermodynamic limit, in thermal equilibrium at
temperature $T$) that the spins $\sigma_i$ and $\sigma_j$ at the sites $i$ and
$j$ in the lattice have the value $a \in I_q$.  At $T=\infty$, all spin
configurations occur with equal probability, so the probability that $\sigma_i$
has a particular value $a$ is just $1/q$, and similarly with $\sigma_j$, so
$P_{aa}(i,j)=1/q^2$ at $T=\infty$. To define a correlation function with the
usual property that in the high-temperature phase, as the distance $r$ between
the spins goes to infinity, they should be completely uncorrelated, one must
therefore subtract this $1/q^2$ term. That is, in the Potts model, one defines
the spin-spin correlation function as (e.g., \cite{wurev})
\beq
\Gamma_{aa}(i,j) = P_{aa}(i,j) - \frac{1}{q^2} \ . 
\label{gamma_cor}
\eeq
Thus, by construction, $\Gamma_{aa}(i,j)=0$ at $T=\infty$. 
At $T=0$, in the ferromagnetic Potts model, all of the spins take on the same
value in the set $I_q$.  Let us say that an infinitesimally small 
external field has been applied to favor the value $a \in I_q$, so then 
$P_{aa}(i,j)=1/q$ and hence, under these conditions, 
\beq
\Gamma_{aa}(i,j) = \frac{1}{q} - \frac{1}{q^2} = \frac{q-1}{q^2} \quad 
{\rm at} \ T=0 \ . 
\label{gamma_zero_temp}
\eeq
In the ferromagnetic Potts model with a conventional external (uniform)
magnetic field favoring a single spin model, the magnetic order parameter,
${\cal M}$, normalized so that it is unity at $T=0$, is then related to this
spin-spin correlation function $\Gamma_{aa}(i,j)$ according to
\beq
{\cal M}
 = \Bigg ( \frac{q^2}{q-1}\Bigg ) \, \lim_{r \to \infty} \Gamma_{aa}(i,j) \ .
\label{Ms1_corfun_rel}
\eeq

Although the quantity $-\partial F/\partial H = \partial f/\partial h$ yields
one measure of magnetic ordering, it is not the order parameter itself, in
contrast to the situation with both the Ising and O($N$) spin models. Instead,
as is evident from its definition, this partial derivative is equal to the
fraction of the total number of sites in the (thermodynamic limit of the)
lattice with spins taking one particular value out of the set of $q$
values. We denote this as
\beq
M = - \frac{\partial F}{\partial H} = \frac{\partial f}{\partial h} 
= w\frac{\partial f}{\partial w} \ . 
\label{M}
\eeq
At $T=\infty$, since all spin values are weighted equally, it follows that 
the fraction of the spins taking on any particular value is $1/q$, i.e., 
\beq
M = \frac{1}{q} \quad {\rm at} \ T=\infty \ . 
\label{Ms1_infinite_temp}
\eeq
In the opposite limit of zero temperature, given that $J > 0$, the spin-spin
interaction forces all of the spins to have the same value. There is then a
dichotomy in the behavior of the system, depending on whether $H$ is positive
or negative.  If $H > 0$, then the external field forces this spin value 
to be the single value in $I_s$ favored by this field, so 
\beq
M = 1 \quad {\rm at} \ T=0 \ {\rm if} \ H > 0 \ ,
\label{Ms1_zero_temp_H_positive}
\eeq
If, on the other hand, $H < 0$, then the single spin value that all the spins 
are forced to have by the spin-spin interaction to lie in the orthogonal 
complement, which, for this case is the set $I_s^\perp=\{2,...,q\}$, 
so the fraction in $I_s$ is zero. Hence, 
\beq
M = 0 \quad {\rm at} \ T=0 \ {\rm if} \ H < 0 \ . 
\label{Ms1_zero_temp_H_negative}
\eeq
The zero-field values of $M$ at a given temperature for the respective 
cases $H > 0$ and $H < 0$ are $M_{0^+} = \lim_{H \to 0^+}M$ and 
$M_{0^-} = \lim_{H \to 0^-}M$.

Because the zero-field value of $M$ does not vanish in the high-temperatures,
$S_q$-symmetric phase, it cannot be the order parameter of the model, but
instead is an auxiliary measure of the magnetic ordering per site. In the
literature, studies focused on the case $H > 0$ in formulating an appropriate
order parameter. In this case, to define an order parameter, one subtracts the
$T=\infty$ value of $M$ from the value for general $T$ and normalizes the
result so that the order parameter saturates at unity at $T=0$. This yields 
a measure of spin ordering that we denote as ${\cal M}$:
\beq
{\cal M} = \frac{M-M_{T=\infty}}
 {M_{T=0} - M_{T=\infty}} = 
 \frac{M-\frac{1}{q}} {1 - \frac{1}{q}} =  \frac{qM-1}{q-1} \ . 
\label{Ms1_Mcal_rel}
\eeq
The zero-field value of this order parameter, i.e., the 
spontaneous magnetization, is then 
\beq
{\cal M}_0 = \lim_{H \to 0} {\cal M} \ . 
\label{Mcals1_0}
\eeq
This construction was given for the specific case $q=3$ in
\cite{straley_fisher} and for general $q$ in \cite{binder} (see also
\cite{wurev}, where ${\cal M}_0$ was denoted as $m_0$).  Series expansions
\cite{straley_fisher} and Monte-Carlo simulations \cite{binder} for the
two-dimensional Potts model were consistent with the behavior expected of an
order parameter, namely ${\cal M}_0=0$ in the high-temperature $S_q$-symmetric
phase and ${\cal M} > 0$ in the low-temperature phase with spontaneous symmetry
breaking of the global $S_q$ symmetry to $S_{q-1}$.

A parenthetical remark is in order concerning the trivial case $q=1$ where the
spins are all frozen to have the same value and hence are nondynamical.  In
this $q=1$ case, up to a prefactor $w^n$, the partition function is equal to
the free-field result, $w^nZ(G,q,v)$.  As a consequence, for any temperature,
$M=1$, so ${\cal M}$ has an indeterminate form $0/0$.  Thus, in using the
formula (\ref{Ms1_Mcal_rel}), one restricts to the range $q \ge 2$.


\section{Measures of Spin Ordering in the Potts Model with Generalized
Magnetic Field}
\label{measures_section}

In this section we present our new results on measures of spin ordering in the
Potts model with a generalized magnetic field, including, in particular, the
order parameter for this model.  In the limit $T \to \infty$, $P_{aa}(i,j) =
1/q^2$, independent of $s$. This is a consequence of the fact that the
Boltzmann weight $e^{-\beta {\cal H}}$ in the expression for $Z$ reduces to 1
for $\beta=0$, and so the spins are completely random.
However, the auxiliary measure of spin ordering, $M$, behaves
differently in the Potts model with a generalized versus conventional
external magnetic field. From the basic definition, calculating $M$ from
Eq. (\ref{M}) and then letting $T \to \infty$, we find the following
general behavior:
\beq
M = \frac{s}{q} \quad {\rm at} \ T=\infty \
{\rm and \ any \ finite} \ H \  \ .
\label{Mgen_infinite_temp}
\eeq
In the opposite limit, $T \to 0$, the value of $M$ again depends on
the sign of $H$. If $H > 0$, then (given that $J > 0$), the spin-spin
interaction forces all spins to have the same value, and the presence
of the external field forces this value to lie in the set $I_s$, so
\beq
\lim_{T \to 0} M = 1 \ {\rm for} \ H > 0 \ .
\label{Mgen_zero_temp_H_positive}
\eeq
In this $T \to 0$ limit (again, given that $J > 0$), if $H < 0$, then 
the spin-spin interaction forces all spins to have the same value and this
value lies in the orthogonal complement $I_s^\perp$, so
the fraction of spins in $I_s$ is 0:
\beq
\lim_{T \to 0} M = 0 \ {\rm for} \ H < 0 \ .
\label{Mgen_zero_temp_H_negative}
\eeq
Finally, we record the behavior of $M$ in the limits $H\to \pm
\infty$ at fixed finite nonzero temperature. In terms of the Boltzmann weights,
these two limits are $w \to \infty$ and $w \to 0$ with $v$ finite. 
As $H \to \infty$ in this limit, all of the spins must 
take on values in $I_s$, so
\beq
\lim_{H \to \infty} M = 1 \ {\rm for \ any \ finite \ nonzero} \ T \ .
\label{Mgen_H_infinity}
\eeq
If $H \to -\infty$, then all spins must take on values in $I_s^\perp$, so the
fraction in $I_s$ is zero for any temperature including $T=0$:
\beq
\lim_{H \to -\infty} M = 0 \quad {\rm for \ any \ finite \ nonzero} \ T  \ .
\label{Mgen_H_minus_infinity}
\eeq
In order to obtain the zero-field value
of $M$ at a given temperature, as in the case of a conventional magnetic
field, one would calculate $Z(G,q,s,v,w)$ on a given lattice graph $G$, take
the thermodynamic limit, then calculate $M$ and take the limit
$H \to 0^+$ or $H \to 0^-$.

We now construct a magnetic order parameter ${\cal M}$ for the Potts model in
a generalized magnetic field. As noted above, given the identity (\ref{zsym}),
we can, without loss of generality, restrict to $H > 0$ and we shall do so
henceforth.  We obtain
\beq
{\cal M} = \frac{M - M_{T=\infty}}
 {M_{T=0} - M_{T=\infty}} =
 \frac{M-\frac{s}{q}} {1 - \frac{s}{q}} = \frac{qM-s}{q-s} \ .
\label{Mgen_Mcal_rel}
\eeq
The spontaneous magnetization is then 
\beq
{\cal M}_0 = \lim_{H \to 0^+} {\cal M} \ .
\label{Mgen_0}
\eeq
A word is in order concerning the apparent pole at $s=q$.  If $s=q$,
then the presence of the external field simply adds a constant term
$-Hn$ to ${\cal H}$, or equivalently, i.e., the partition function is
equivalent to the product of the factor $w^n$ times the zero-field
$Z$, as specified in the identity (\ref{zwsq}), so that
\beq
M=1 \quad {\rm if} \ s=q \ , \label{M_gen_s_equal_q}
\eeq
independent of temperature. Hence, just as was the case with the expression
(\ref{Ms1_Mcal_rel}) for a conventional magnetic field favoring just one spin,
so also here, the expression (\ref{Mgen_Mcal_rel}) takes the indeterminate form
$0/0$ in this case.  Hence, in using (\ref{Mgen_Mcal_rel}), we restrict $s$ to
the interval $1 \le s \le q-1$.


\section{Some Explicit Examples}
\label{example_section}

Some explicit examples illustrate the use of (\ref{Mgen_Mcal_rel}) for the
order parameter. Although a Peierls-type argument shows that there is no
spontaneous symmetry breaking of the $S_s \otimes S_{q-s}$ symmetry
(\ref{symgroup}) on (the $n \to \infty$ limit of a) one-dimensional lattice or
quasi-one-dimensional lattice strip, these
types of lattices are, nevertheless, useful to illustrate some features of
Eq. (\ref{Mgen_Mcal_rel}).  

\subsection{1D Lattice}

As a first example, we use the exact expression for $Z(G,q,s,v,w)$ on a
one-dimension lattice derived in \cite{phs}.  This yields the reduced
dimensionless free energy per site (in the thermodynamic limit, in the notation
of \cite{phs})
\beq
f(1D,q,s,v,w) = \ln( \lambda_{Z,1,0,1}) \ ,
\label{f1d}
\eeq
where
\beq
\lambda_{Z,1,0,1} = \frac{1}{2}\Big ( A + \sqrt{R} \ \Big ) \ , 
\label{lambda_cn}
\eeq
with 
\beq
A = q+s(w-1)+v(w+1)
\label{a}
\eeq
and 
\beq
R = A^2-4v(q+v)w \ . 
\label{R}
\eeq
(As expected, in the thermodynamic limit, this result applies
independent of the boundary conditions.)  The resultant auxiliary
measure of spin ordering, $M$, is 
\beq
M = \frac{w}{\sqrt{R}} \, \Bigg [ s+v - \frac{2v(q+v)}{A+\sqrt{R}} \, \Bigg ]
\ . 
\label{Mgen_1D}
\eeq
It is straightforward to confirm that Eq. (\ref{Mgen_1D}) satisfies
the general relations (\ref{Mgen_infinite_temp})-(\ref{Mgen_H_minus_infinity}).
As $T \to \infty$, i.e., $\beta
\to 0$, the variables $v$ and $w$ (for finite $J$ and $H$) approach
the limits $v \to 0$ and $w \to 1$, i.e., $K \to 0$ and $h \to 0$. In
this limit, we calculate a two-variable series expansion of $M$ in $K$
and $h$ and find
\beqs
M &=& \frac{s}{q} \Bigg [ 1 + \bigg (1 -\frac{s}{q} \bigg ) h \Bigg
\{ 1 + \frac{2}{q}K + \frac{1}{2}\Big (1-\frac{2s}{q}\Big )h +
\frac{1}{q}K^2 + \frac{3}{q}\Big (1-\frac{2s}{q}\Big )Kh \cr\cr
&+& \Big (\frac{1}{6}-\frac{s}{q}\Big (1-\frac{s}{q}\Big ) \Big ) h^2
+ O(K^3,K^2h,Kh^2,h^3) \Bigg \} \, \Bigg ] \quad {\rm as} \ T \to \infty \ .
\label{Mgen_high_temp_double_series}
\eeqs
Setting $\beta=0$, one sees that this expansion satisfies the
identity (\ref{Mgen_infinite_temp}).  Substituting
Eq. (\ref{Mgen_high_temp_double_series}) into our general expression for the
order parameter, we obtain
\beqs
{\cal M} &=&
\frac{sh}{q} 
\Bigg [ 1 + \frac{2}{q}K + \frac{1}{2}\Big (1-\frac{2s}{q}\Big )h +
\frac{1}{q}K^2 + \frac{3}{q}\Big (1-\frac{2s}{q}\Big )Kh
+ \Big (\frac{1}{6}-\frac{s}{q}\Big (1-\frac{s}{q}\Big ) \Big ) h^2 \cr\cr
&+& O(K^3,K^2h,Kh^2,h^3) \Bigg \} \, \Bigg ] \quad {\rm as} \ T \to \infty \ , 
\label{Mcal_hkseries}
\eeqs
where the notation $O(K^3,K^2h,Kh^2,h^3)$ refers to terms of order 
$K^3$, $K^2h$, $Kh^2$, or $h^3$ inside the curly brackets.  
The proportionality of ${\cal M}$ to $(s/q)h=(s/q)\beta H$ as 
$\beta \to 0$ is the expression of the Curie-Weiss relation for the induced
magnetization for this model.

Given Eq. (\ref{Mgen_Mcal_rel}) connecting $M$ and 
${\cal M}$, the susceptibilities defined via $M$ and ${\cal M}$ are 
simply related to each other. Defining 
$\chi_M = \partial M/\partial H$ and 
$\chi_{\cal M} = \partial{\cal M}/\partial H$, we have 
\beq
\chi_M = \Big ( 1-\frac{s}{q} \Big ) \, \chi_{\cal M} \ . 
\label{chi_rel}
\eeq
From Eq. (\ref{Mcal_hkseries}), it follows that 
the two-variable high-temperature Taylor series expansion of $\chi_{\cal M}$ 
in powers of $K$ and $h$ is given by 
\beqs
\beta^{-1} \chi_{\cal M} &=&  \frac{s}{q} 
\Bigg [ 1 + \frac{2}{q}K + \Big (1-\frac{2s}{q}\Big )h +
\frac{1}{q}K^2 + \frac{6}{q}\Big (1-\frac{2s}{q}\Big )Kh
+ \Big (\frac{1}{2}-\frac{3s}{q}\Big (1-\frac{s}{q}\Big ) \Big ) h^2 \cr\cr
&+& O(K^3,K^2h,Kh^2,h^3) \Bigg \} \, \Bigg ] \quad {\rm as} \ T \to \infty \ . 
\label{chi_Mcal_hkseries}
\eeqs

For $H \to \infty$ at finite $T$ (equivalently, 
$w \to \infty$ with finite $v$), we calculate the Taylor series expansion
\beq
M = 1 - \frac{s(q-s)}{(s+v)^2w} + \frac{s(q-s)[s(q-s)-2v(q+v)]}{(s+v)^4w^2}
+ O\Big ( \frac{1}{w^3} \Big )
\label{Mgen_largeHseries}
\eeq
and hence
\beq
{\cal M} = 1 - \frac{sq}{(s+v)^2w} + \frac{sq[s(q-s)-2v(q+v)]}{(s+v)^4w^2}
+ O\Big ( \frac{1}{w^3} \Big ) \ .
\label{Mcalgen_largeHseries}
\eeq

\begin{figure}
  \begin{center}
    \includegraphics[height=0.7\linewidth]{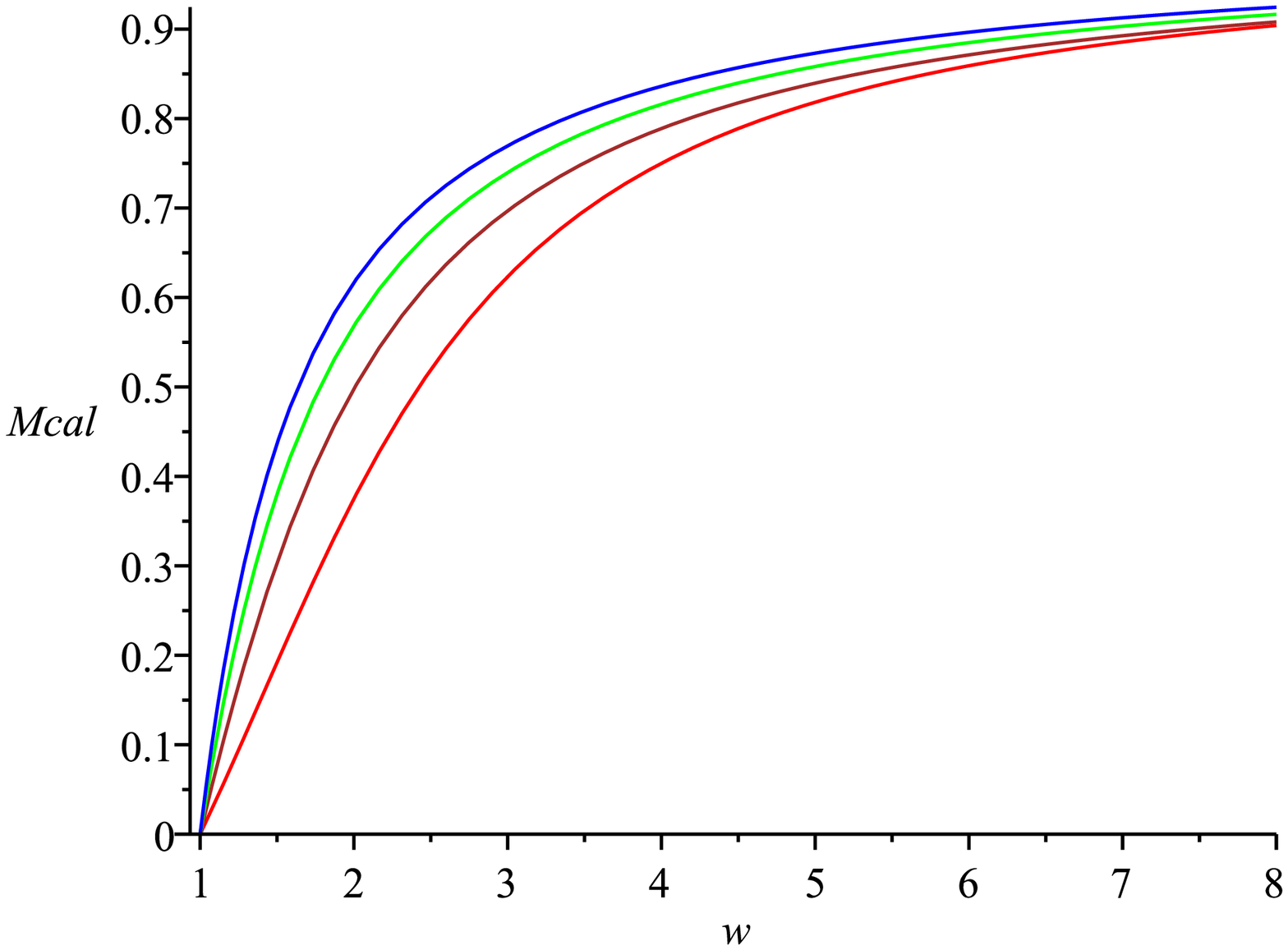}
  \end{center}
  \caption{Plot of ${\cal M}$ (denoted Mcal in this and other figures)
    for the 1D lattice as a function of $w$ in
  the range $1 \le w \le 8$ for the illustrative values $q=5$ and
  $v=2$.  For $w \gsim 1$, the curves, going from bottom to top, refer
  to $s=1, \ 2, \ 3, \ 4$. The colors are $s=1$ (red), $s=2$ (brown),
$s=3$ (green), and $s=4$ (blue).} 
\label{Mcal_w1to8_fig}
\end{figure}

\begin{figure}
  \begin{center}
    \includegraphics[height=0.7\linewidth]{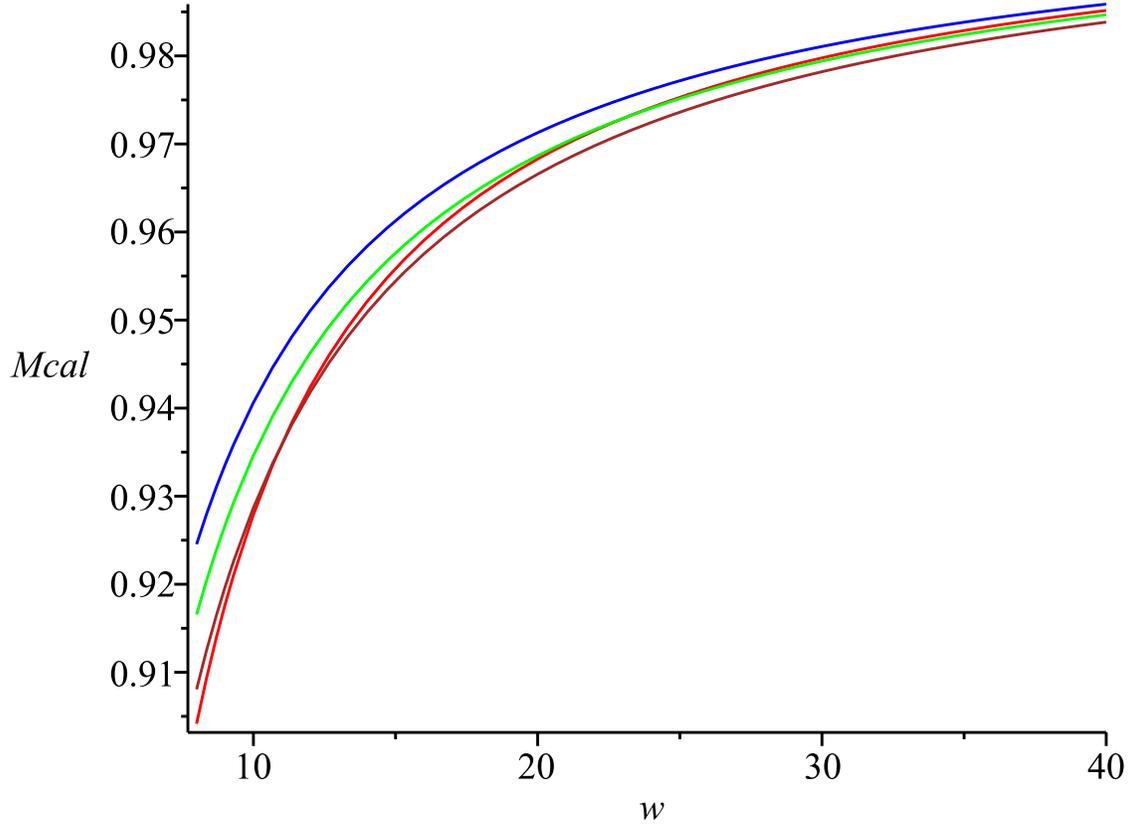}
  \end{center}
\caption{Plot of ${\cal M}$ for the 1D lattice as a function of $w$ in
  the range $8 \le w \le 40$ for the illustrative values $q=5$ and
  $v=2$.  For $w=8$, the curves, going from bottom to top, refer to
  $s=1, \ 2, \ 3, \ 4$; crossimgs of curves are evident for larger $w$
  values.  The color coding is the same as in Fig. \ref{Mcal_w1to8_fig}.}
\label{Mcal_w8to40_fig}
\end{figure}

To show these results numerically for a typical case, we take the illustrative
values $q=5$ and $v=2$. In Figs. \ref{Mcal_w1to8_fig} and
\ref{Mcal_w8to40_fig} we plot $\cal M$ for this 1D lattice
as a function of $w$ in the intervals $1 \le w \le 8$ and $8 \le w \le 40$.
Fixing the value of $v$
corresponds most simply to fixing the values of $J$ and $T$, so that
the variation in $w$ then amounts to a variation in $H$ at fixed $T$.
The results show that, as expected, $\cal M$ increases monotonically
with increasing $w$ and thus $H$, for fixed $T$. For small $h$, i.e.,
$w-1 \to 0^+$, the values of ${\cal M}$ satisfy the
relation ${\cal M} = (s/q)h$, in accord with (\ref{Mcal_hkseries}) and
hence are larger for larger $s$. However, as is evident from the
series expansion (\ref{Mcalgen_largeHseries}) and from Fig.
\ref{Mcal_w8to40_fig}, this monotonicity is not preserved in ${\cal M}$
for this 1D lattice at large $w$.

\subsection{$L_y=2$ Lattice Strips}

\begin{figure}
  \begin{center}
    \includegraphics[height=0.7\linewidth]{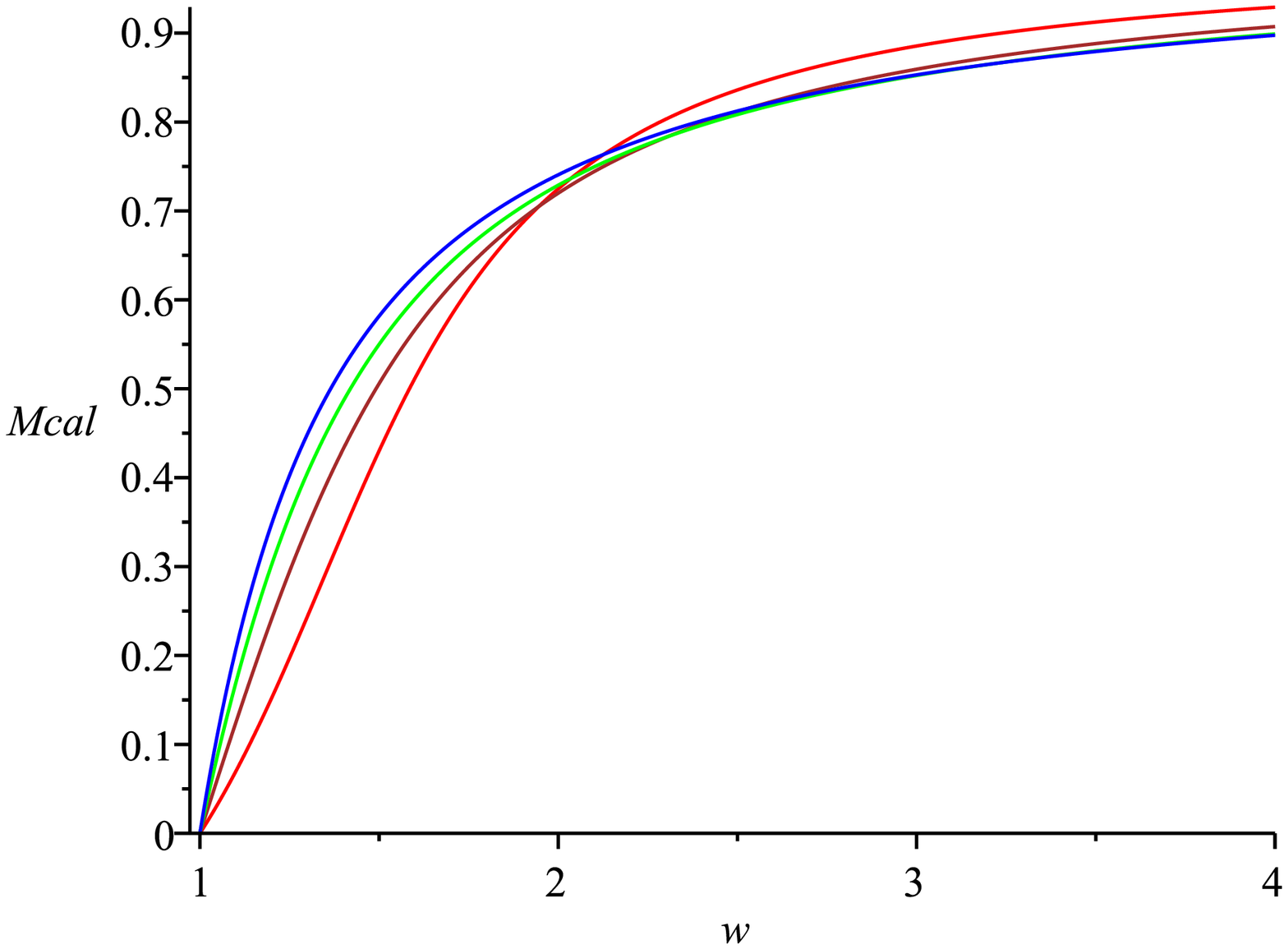}
  \end{center}
  \caption{Plot of ${\cal M}$ for the infinite-length
    strip of the square lattice of width $L_y=2$,
as a function of $w$ in the range $1 \le w \le 4$
for the illustrative values $q=5$ and $v=2$.
For $w \gsim 1$, the curves, going from bottom to top, refer
to $s=1, \ 2, \ 3, \ 4$.
The color coding is the same as in Fig. \ref{Mcal_w1to8_fig}.}
    \label{Mcalsq_w1to4_fig}
\end{figure}

\begin{figure}
  \begin{center}
    \includegraphics[height=0.7\linewidth]{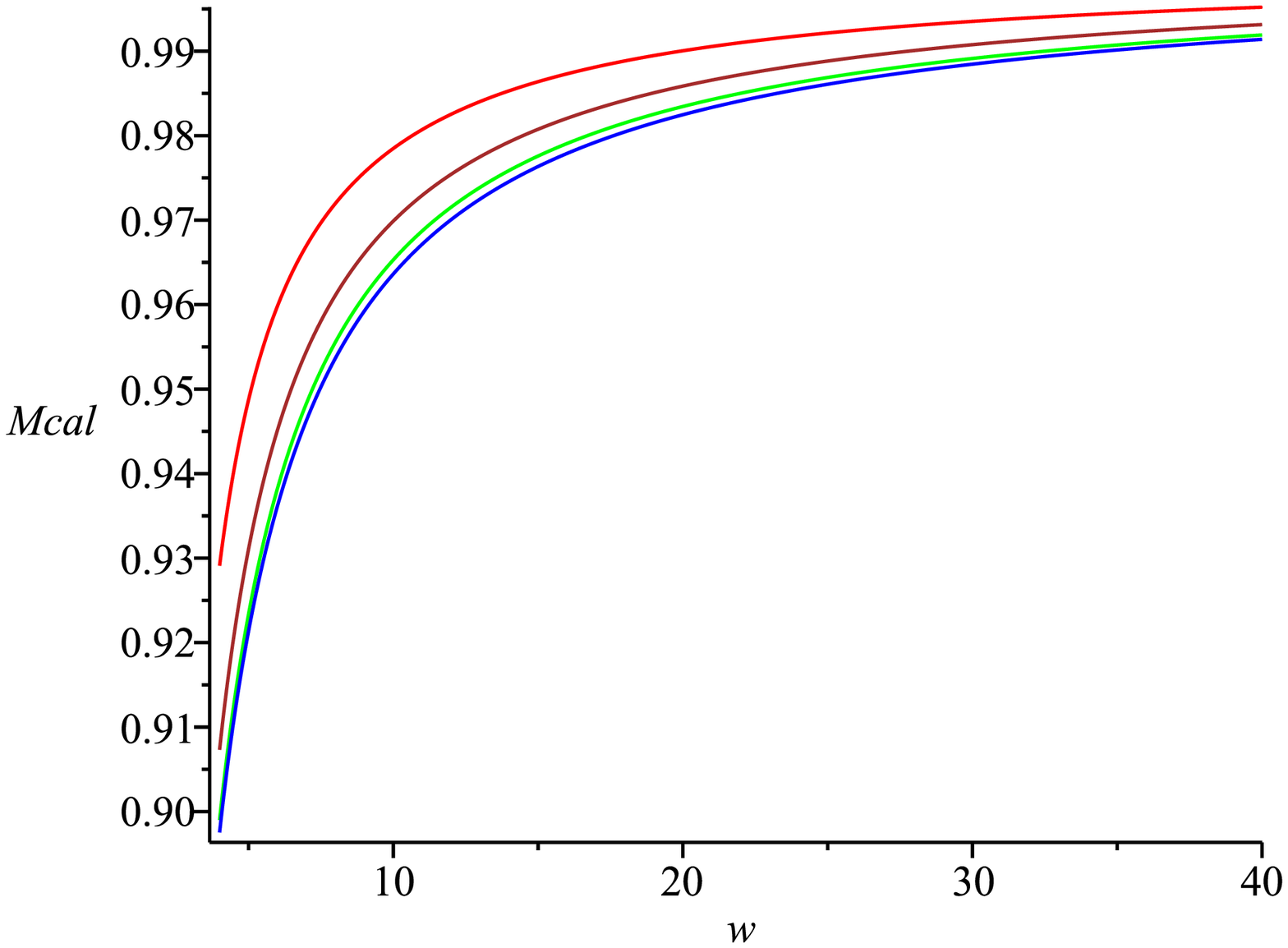}
  \end{center}
  \caption{Plot of ${\cal M}$ for the infinite-length
    strip of the square lattice of width $L_y=2$,
as a function of $w$ in the range $4 \le w \le 40$
  for the illustrative values $q=5$ and $v=2$. The curves,   
  going from bottom to top, refer to $s=4, \ 3, \ 2, \ 1$.
The color coding is the same as in Fig. \ref{Mcal_w1to8_fig}.}
\label{Mcalsq_w4to40_fig}
\end{figure}

\begin{figure}
  \begin{center}
    \includegraphics[height=0.7\linewidth]{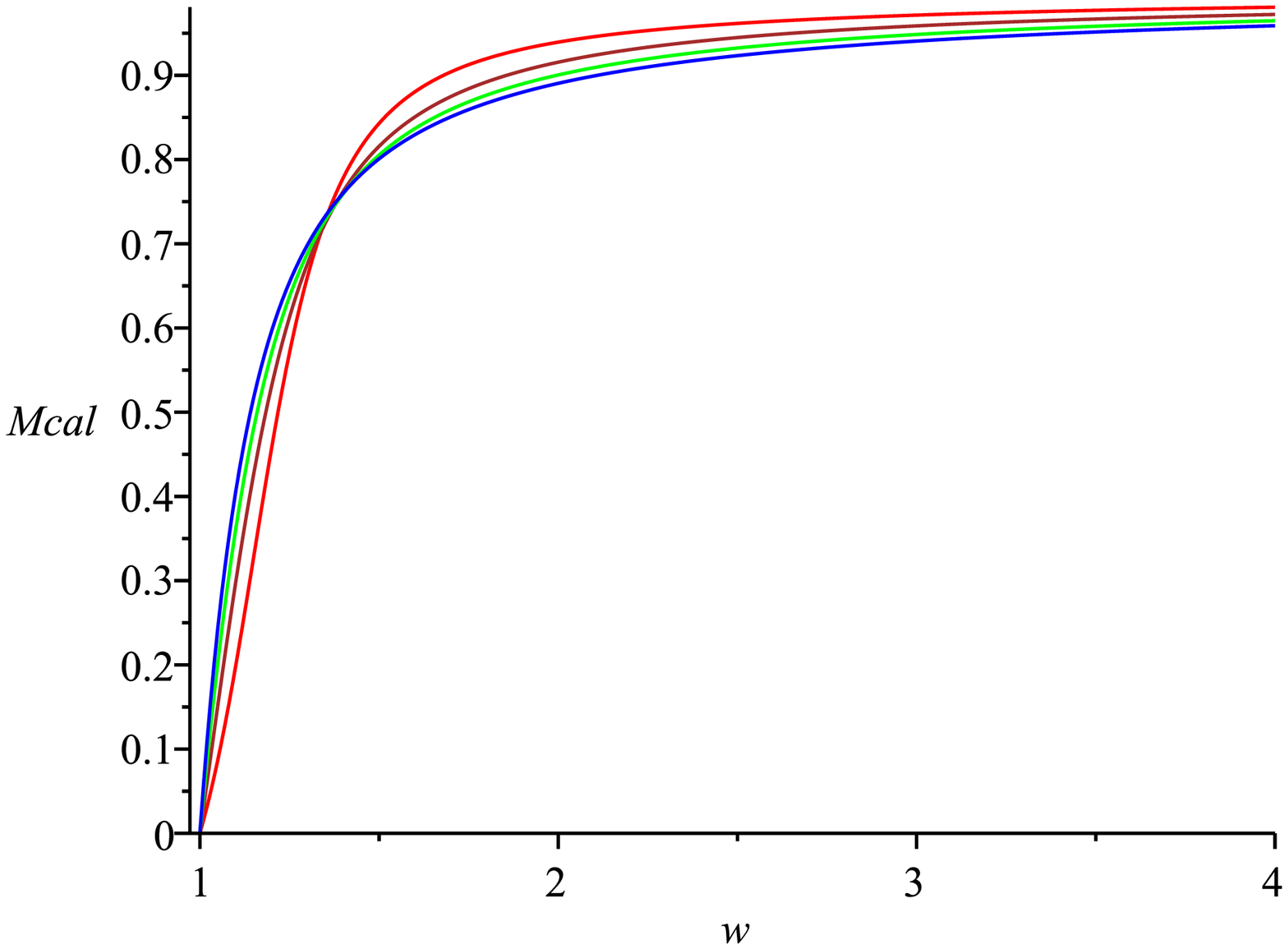}
  \end{center}
  \caption{Plot of ${\cal M}$ for the infinite-length
    strip of the triangular lattice of width $L_y=2$,
as a function of $w$ in the range $1 \le w \le 4$
  for the illustrative values $q=5$ and $v=2$.  For $w \gsim 1$, the curves,
  going from bottom to top, refer to $s=1, \ 2, \ 3, \ 4$.
  The color coding is the same as in Fig. \ref{Mcal_w1to8_fig}.}
\label{Mcaltri_w1to4_fig}
\end{figure}

\begin{figure}
  \begin{center}
    \includegraphics[height=0.7\linewidth]{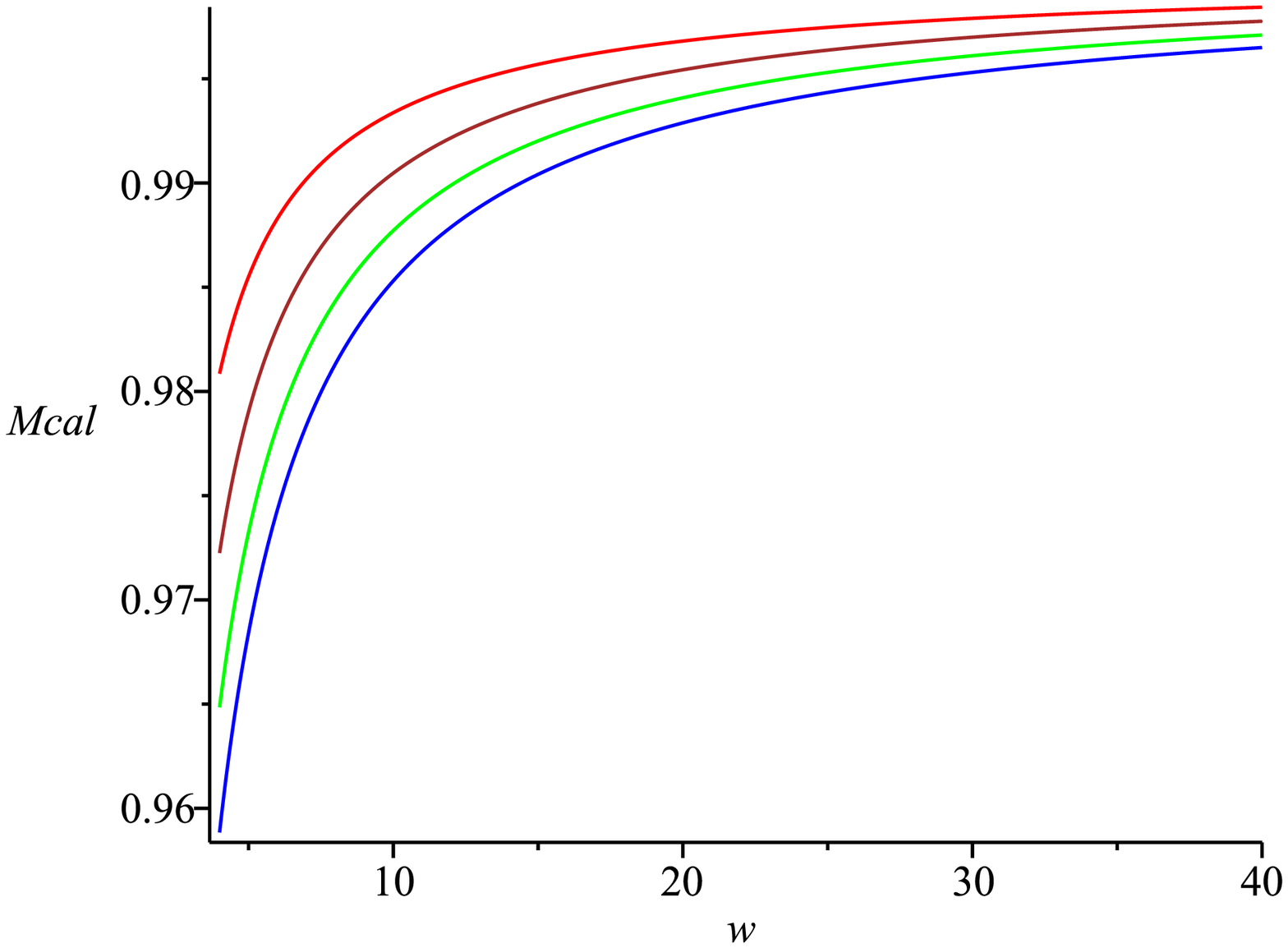}
  \end{center}
  \caption{Plot of ${\cal M}$ for the infinite-length
    strip of the triangular lattice of width $L_y=2$,
as a function of $w$ in the range $4 \le w \le 40$
  for the illustrative values $q=5$ and $v=2$.  The curves,
  going from bottom to top, refer to $s=4, \ 3, \ 2, \ 1$.
  The color coding is the same as in Fig. \ref{Mcal_w1to8_fig}.}
\label{Mcaltri_w4to40_fig}
\end{figure}

In \cite{zhlad} we calculated $Z(G,q,s,v,w)$ for the width $L_y$ strips of the
square and triangular lattice. 
This work generalized our previous calculations of
$Z(G,q,v)$ in zero external field on these strips \cite{a,ta}. 
In the infinite-length limit 
(independent of longitudinal boundary conditions), the reduced free
energy is given, respectively, by 
$f_{sq,L_y=2} = (1/2)\ln \lambda_{sq,L_y=2}$ and 
$f_{tri,L_y=2} = (1/2)\ln \lambda_{tri,L_y=2}$, where $\lambda_{sq,L_y=2}$ and 
$\lambda_{tri,L_y=2}$ are roots of respective degree-5 and degree-6 algebraic
equations. Hence, it is not possible to calculate the derivatives 
$M = w \partial f/\partial w$ analytically to give explicit expressions 
for $M$ and ${\cal M}$ for these infinite-length lattice strips.  However, 
using numerical differentiation, it is still possible to  obtain values for 
these quantities, given input values for $v$, $q$, and $s$. Using this
method and again taking the illustrative values $q=5$ and $v=2$, we show 
plots of ${\cal M}$ for the infinite-length strips of the square and
triangular lattices with width $L_y=2$  
in Figs. \ref{Mcalsq_w1to4_fig}-\ref{Mcaltri_w4to40_fig}. 
As was the case with the 1D lattice, for small $h$, 
the values of ${\cal M}$ satisfy the
relation ${\cal M} = (s/q)h$, and thus are larger for larger $s$, but this
monotonicity relation does not apply for large $w$.  


\section{Thermodynamic Properties and Critical Behavior} 
\label{critical_section}

As discussed in Sect. \ref{gen_section} above, 
in the presence of the generalized external magnetic field defined in
Eq. (\ref{hp}), the symmetry group of ${\cal H}$ and $Z$ is reduced from 
${\cal S}_q$
at $H=0$ to the tensor product in Eq. (\ref{symgroup}), and this further
simplifies to ${\cal S}_{q-1}$ if $s=1$ in which case, the 
external field favors or disfavors only a single spin value. From the
identity (\ref{zsym}), the case $s=q-1$ is effectively equivalent to the 
conventional case $s=1$.  However, if $s$ is in the interval
\beq
2 \le s \le q-2 \ ,
\label{srange}
\eeq
then the general model of Eqs. (\ref{ham}) and (\ref{hp}) exhibits
properties that are interestingly different from those of a $q$-state
Potts model in a conventional magnetic field.  In this section we will
consider both signs of $H$ and $J$.  With a conventional magnetic
field, at a given temperature $T$, if $H \gg |J|$, the interaction
with the external field dominates over the spin-spin interaction, and
if $h = \beta H$ is sufficiently large, the spins are frozen to the
single favored value.  In contrast, in the model with a generalized
magnetic field, if $s$ lies in the interval (\ref{srange}), and if
$|H| \gg |J|$, this effectively reduces the model to (i) an $s$-state
Potts model if $H > 0$, or (ii) a $(q-s)$-state Potts model if $H <
0$.  For given values of $q$ and $s$, taking the thermodynamic limit
of a given regular lattice, there are, in general, four types of
possible models, depending on the sign of $H$ and the sign of $J$.  A
discussion of these models, including the types of critical behavior,
where present in the case of square lattice, was given in (Section 4
of) \cite{phs2} with details for the illustrative case $q=5$ and
$s=2$. We generalize this here to $q \ge 5$. For $H=0$, the
ferromagnetic version of the model has a first-order phase transition,
with spontaneous breaking of the ${\cal S}_5$ symmetry, at $K_c =
\ln(1+\sqrt{q})$, while the antiferromagnetic version has no
finite-temperature phase transition and is disordered even at $T=0$
\cite{wurev,baxterbook}. For $H > 0$ and $H \gg |J|$, the theory
reduces effectively to a two-state Potts model, i.e., an Ising model.
Owing to the bipartite property of the square lattice, there is an
elementary mapping that relates the ferromagnetic and
antiferromagnetic versions of the model, and, as is well known, both
have a second-order phase transition, with spontaneous symmetry
breaking of the ${\cal S}_2 \approx {\mathbb Z}_2$ symmetry, at
$|K_c|=\ln(1+\sqrt{2}) \simeq 0.881$ (where $K=\beta J$), with thermal
and magnetic critical exponents $y_t=1$, $y_h=15/8$, described by the
rational conformal field theory (RCFT) with central charge $c=1/2$.
For $H < 0$ and $|H| \gg |J|$, the theory effectively reduces to a
$(q-2)$-state Potts model.  In the ferromagnetic case, $J > 0$, if (a)
$q=5$, then the resultant 3-state Potts ferromagnet has a
well-understood second-order phase transition, with spontaneous
symmetry breaking of the ${\cal S}_3$ symmetry, at $K_c =
\ln(1+\sqrt{3}) \simeq 1.01$, with thermal and magnetic critical
exponents $y_t=6/5$, $y_h=28/15$, described by a RCFT with central
charge $c=4/5$; (b) if $q=6$, then the resultant 4-state Potts
ferromagnet also has a second-order phase transition with thermal and
magnetic critical exponents $y_t=3/2$, $y_h=15/8$, described by a RCFT
with central charge $c=1$ \cite{wurev,bpz,cft}; and (c) if $q \ge 7$,
then the resultant Potts ferromagnet has a first-order transition
\cite{baxterbook,wurev}.  In the antiferromagnetic case, $J < 0$, if
$q=5$, the resultant 3-state Potts antiferromagnet has no
finite-temperature phase transition but is critical at $T=0$ (without
frustration), with nonzero ground-state entropy per site $S/k_B =
(3/2)\ln(4/3) \simeq 0.432$ \cite{wurev,lieb}. If $q \ge 6$, then the
resultant $(q-2)$-state Potts antiferromagnet (on the square lattice)
does not have any symmetry-breaking phase transition at any finite
temperature and is disordered also at $T=0$. Similar discussions can
be given for other lattices. 


\section{Conclusions}

In this paper we have discussed measures of spin ordering in the
$q$-state Potts model in a generalized external magnetic field that
favors or disfavors spin values in a subset $I_s = \{1,...,s\}$ of the
total set of $q$ values. In particular, we have constructed an order
parameter ${\cal M}$ (given in Eq. (\ref{Mgen_Mcal_rel})) and have
presented an illustrative evaluation of it, together with relevant
series expansions, for the (thermodynamic limit of the)
one-dimensional lattice, as well as quantitative plots of $\cal M$ for
this 1D lattice and for strips of the square and triangular lattices.


\begin{acknowledgments}

  The research of S.-C.C. was supported in part by the Taiwan Ministry of
Science and Technology grant MOST 111-2115-M-006-012-MY2. The research of
  R.S. was supported in part by the U.S. National Science Foundation Grant
  NSF-PHY-22-100533.

\end{acknowledgments}




\end{document}